%% file: puppet.tex
\documentclass[sigchi]{acmart}
\AtBeginDocument{%
  \providecommand\BibTeX{{%
    \normalfont B\kern-0.5em{\scshape i\kern-0.25em b}\kern-0.8em\TeX}}}

\setcopyright{acmcopyright}
\copyrightyear{2018}
\acmYear{2018}
\acmDOI{10.1145/1122445.1122456}

\acmConference[Woodstock '18]{Woodstock '18: ACM Symposium on Neural
  Gaze Detection}{June 03--05, 2018}{Woodstock, NY}
\acmBooktitle{Woodstock '18: ACM Symposium on Neural Gaze Detection,
  June 03--05, 2018, Woodstock, NY}
\acmPrice{15.00}
\acmISBN{978-1-4503-XXXX-X/18/06}

\include{macros}
\begin{document}

\title{PREPRINT: Found Object Puppeteering as a Tool for Rapid Movement Sketching in 3D Animation}

\author{Molly Jane Nicholas}
\email{molecule@berkeley.edu}
\affiliation{
    \institution{University of California, Berkeley}
    \country{USA}
}

\author{Eric Paulos}
\email{paulos@berkeley.edu}
\affiliation{
    \institution{University of California, Berkeley}
    \country{USA}
}

\renewcommand{\shortauthors}{Trovato and Tobin, et al.}

\begin{abstract}
Both expert and novice animators have a need to engage in movement sketching -- low-cost, rapid iteration on a character's movement style -- especially early on in the ideation process. Yet animation tools currently focus on low-level character control mechanisms rather than encouraging engagement with and deep observation of movement. We identify Found Object puppeteering -- where puppeteers manipulate everyday physical objects with their hands -- as a creative practice whose use of material ``jigs'' is uniquely well-positioned to scaffold the novice animator's developing skills. In this paper, we draw on the practice of an expert puppeteer practitioner to inform the design of a system that incorporates physical objects into the animation workflow to scaffold novices into diverse movement exploration while manipulating digital puppets.
\end{abstract}

\ccsdesc[500]{Computer systems organization~Embedded systems}
\ccsdesc[300]{Computer systems organization~Redundancy}
\ccsdesc{Computer systems organization~Robotics}
\ccsdesc[100]{Networks~Network reliability}

\keywords{animation; puppetry; creativity support tool}

\begin{teaserfigure}
  \includegraphics[width=\textwidth]{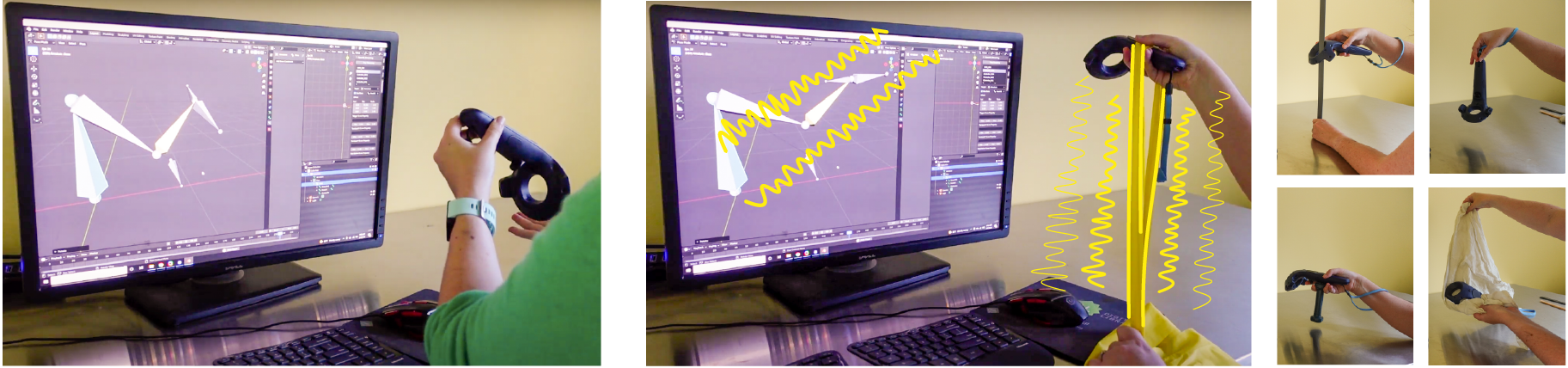}
  \caption{Incorporating physical materials -- such as a stretchy band, weights, fabric -- into the movement sketching process has the potential to enable novice animators to deeply engage with movement qualities. Left: Tangible controllers allow for embodied manipulation of 3D digital models. Center: ``Found objects'' such as a stretchy band act as material jigs. Recording the movement of a digital character while pulling the band adds an ineffable quality of tension to the movement. More importantly, playing with physical materials enhances the ideation process for novice animators. Right: Novice animators using the tool for the first time experimented with using material jigs in diverse ways. Clockwise from upper left: Using a stick to constrain movement along a path; dangling the controller to allow gravity to generate novel movements; dropping the controller into a piece of fabric; and holding a set of weights to embody the experience of a `heavy' or `sad' character. \katherine{this might be because I'm not incredibly familiar with what the HTC vive controller looks like, but I have a hard time understanding what's happening in anything except the leftmost figure. Without reading the rest of the paper, I also can maybe get a vague sense of a "jig" from the stretchy band (though I don't know what the yellow squigglies are on the screen?) and the fabric, but the plastic bar and hand-held weight insets break my brain a little bit} \sarah{squiggles are confusing.  "tangible controller" is a weird phrase to start with.  "support diverse movement prototyping" is vague}} \rawn{squiggles made it hard to see the band - maybe just some arrows showing the motion?}
  \Description{}
  \label{fig:teaser}
\end{teaserfigure}

\maketitle

\section{Introduction}
In animation, movement is key for conveying a character's personality, emotions, story, and meaning. However, current animation tools for designing character movement remain challenging to learn, requiring extensive investment in time and effort~\cite{dontcheva2003layered}
limiting both their adoption by novice animators, and their usefulness in early ideation sketching even for experts.
Additionally, the use of techniques such as creating keyframes and interpolating between them keeps designers focused on low-level mechanisms rather than allowing them to quickly engage in 
sketching behaviors -- early ideation that is quick, exploratory, ambiguous, gestural~\cite{buxtonSketching, buxton2010sketching} -- a key step in the creative process~\cite{buxton2010sketching, kaufman2016creativity}. Sketching can be understood as a low-cost design strategy that allows experienced sketchers to engage in a reflection-in-action constructionist process~\cite{goldschmidt2014modeling}, or as early externalizations of an idea, or as filters and manifestations of design ideas~\cite{lim2008anatomy}. As such, the early ideation sketching process benefits from tools that provide a low \textit{threshold}~\cite{myers2000past} and \textit{paths of least resistance}~\cite{myers2000past} to expressive behaviours. In this work, we ask the question: \textit{How can animation tools scaffold animators towards beneficial movement sketching techniques?}

\begin{figure}[t]
    \centering
    \includegraphics[width=.52\columnwidth]{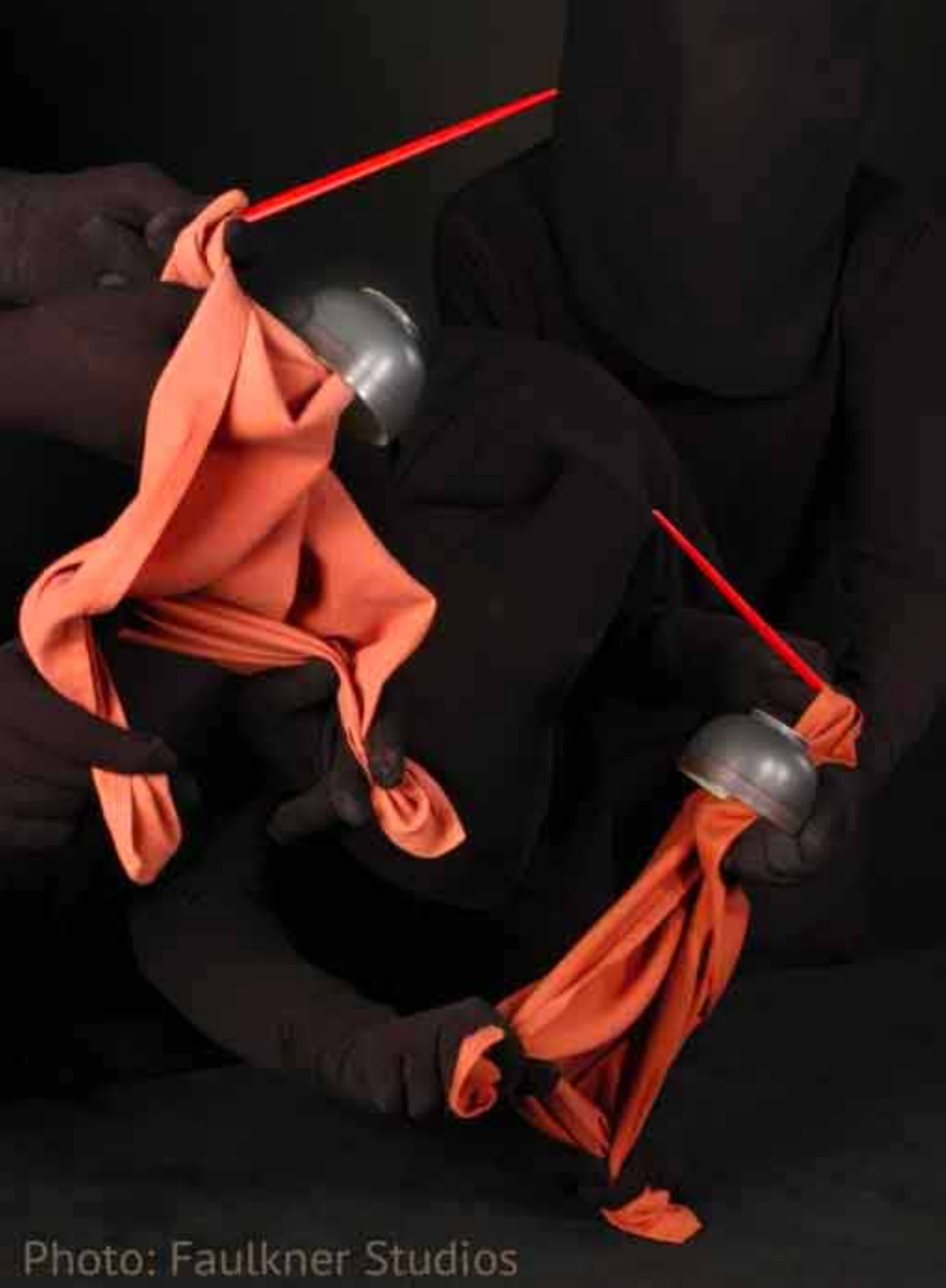}
    \caption{Found Object puppeteering, also known as ``live 3D animation'' involves puppeteers manipulating everyday materials with their hands. In this image, puppeteers wearing all black manipulate napkins, a tea cup, and chopsticks to create two sword-fighting characters.}
    \label{fig:lun-fan}
\end{figure}

\begin{figure*}[t]
    \centering
    \includegraphics[width=.92\linewidth]{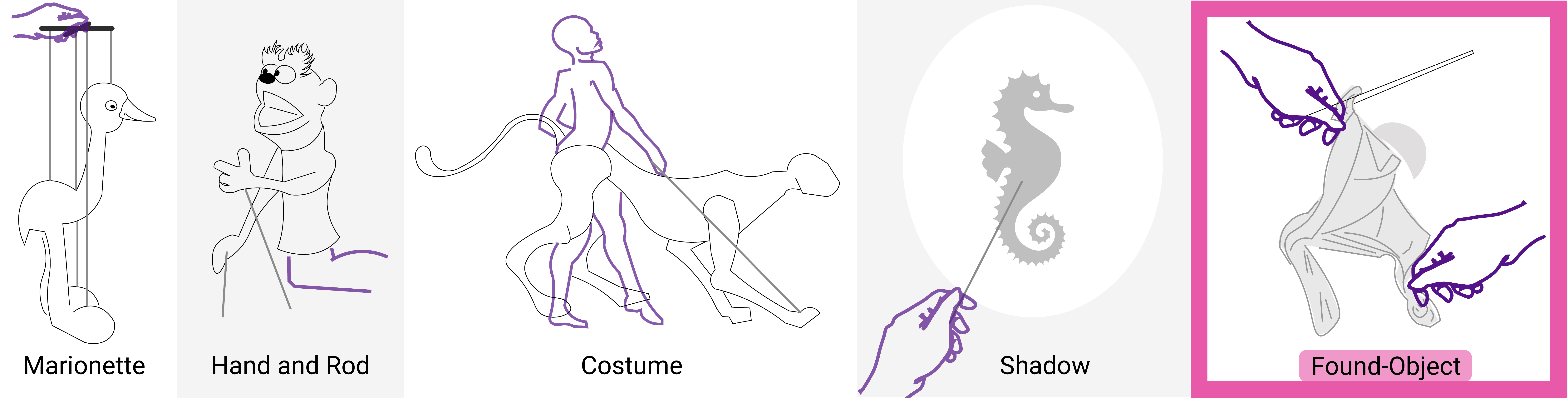}
    \caption{There are many different forms of puppeteering, of which the majority are not as potentially beneficial to incorporate into the animator's workflow. From left to right, Marionettes like Pinocchio are suspended from strings attached to a hand-held control mechanism. Hand and Rod puppets (e.g. ``the Muppets'') are controlled with a hand inside the head opening and closing the mouth, and rods attached to both hands. Costume puppets like Big Bird, or many of the creatures in Julie Taymor's The Lion King on Broadway, incorporate the puppeteer's body into the character. Shadow puppets rely on light and can either be made of cut-outs (as in traditional Indonesian Wayang Kulit~\cite{escobar2017tangible}), or with hand shapes. Found Object puppetry involves the manipulation of materials such as napkins, paper bags, plastic forks, etc. Nearly any object can become a puppet in this style of puppetry. Such a bricolage practice~\cite{vallgaarda2015interaction} is uniquely positioned to contribute character design strategies to the world of animation via material jigs that can define, shape, and influence movement qualities of digital puppets.}
    \label{fig:forms-of-puppeteering}
\end{figure*}

Rapid embodied sketching is the domain of another creative field: Puppeteers have been bringing inanimate objects to life to the delight of many for hundreds of years. In Found Object puppetry, everyday materials directly shape the design of compelling characters. For example, a napkin may be crumpled in a particular way and combined with a ceramic cup and a stick to create a fighting character (see Figure \ref{fig:lun-fan}). Also called ``live 3D animation'', this puppeteering technique is a \textit{bricolage} practice~\cite{vallgaarda2015interaction} that relies on a ``knowing-through-action''~\cite{Dalsgaard2014, PeterDalsgaard2017}, reflective conversation with materials~\cite{schon1979reflective}. Because it relies on physically manipulating ``objects at hand'' (rather than requiring a constructed puppet, as in Marionette, Hand and Rod, Costume, or Shadow Puppetry (see Figure \ref{fig:forms-of-puppeteering}), this technique is particularly well-suited to supporting quick engagement in embodied movement exploration. 

Specifically, we suggest that the materials typically used by Found Object puppeteers could helpfully influence animation techniques, if they could be incorporated into the animator's workflow. In this paper, we introduce a system for novice digital animators which incorporates aspects of analog, tangible, Found Object puppeteering. The expert Puppeteer we interviewed described the materials that he uses in his practice as helpfully constraining his movements, much like the way that jigs and fixtures support woodworkers by providing selective constraints to motion. While jigs in woodworking are typically solid and hold cutting materials securely in place, and the Puppeteer used soft or flexible materials such as a napkin or a jacket, the core idea of an external tool that helpfully limits movement remains the same. We therefore also refer to the materials we incorporate into the animation process as \textit{jigs}. We show that by conceptualizing these ``found objects'' as \textit{material jigs} and incorporating them into the animator workflow, novices use the materials to engage in embodied exploration to generate movement sketches: using a stretchy band between both arms to create tense, vibratory movement or hanging a controller by a piece of fabric to capture naturalistic pendular effects. Together, tangible animation controllers and material jigs enhance the novice animator's character design practice and ability to engage in a reflective, embodied conversation with both the digital sketching output and the physical sketching materials themselves~\cite{Schon1992, klemmer2006bodies}.

In this paper, we draw on techniques from expert puppeteer practitioners to inform the design of a system that allows novice animators to engage in embodied movement sketching practices. We identify Found Object-style puppeteering as uniquely positioned to contribute character design strategies to the world of animation. We first describe strategies and techniques used by expert practitioners in two distinct but related fields -- animation and puppeteering -- and describe how the Found Object puppeteering strategy of material ``jigs'' can be fruitfully imported into the core animation workflow. Next, we describe our authoring tool, \systemTwo, which allows designers to define, layer, edit, and replay motion-tracked character animations via the manipulation of tangible controllers and material jigs. We then share the results of an exploratory evaluation with participants experiencing the tool for the first time. Finally, we discuss how this concept of jigs applies to the world of digital animation, and suggest future directions for exploration.

\section{Related Work}

\subsection{Tangible Tools and Systems}
Tangible interfaces have long been recognized for providing benefits in contexts that require experimentation, muscle memory, tacit learning, and the ambiguity and complexity of the physical world~\cite{klemmer2006bodies}. Analog tangible interfaces -- such as clay -- are known for enabling a rich ``conversation with materials''~\cite{schon1979reflective} which designers frequently seek to recreate with digital materials~\cite{moradi2022glaze, torres2019hybrid, torres2019conversation}. For example, Jones et al. created a system that enables designers to fabricate clay sculptures~\cite{Jones:2016:YSY:2858036.2858493}, arguing that an interactive, physical prototype affords a more accurate, iterative, and responsive design process.
Raffles et al. introduced Topobo, a tangible interface designed to support children learning about how ``balance, leverage and gravity affect moving structures''~\cite{raffle2004topobo}. Tangible systems also enable the capture of physical performances for archive purposes~\cite{escobar2017tangible}. ChronoFab is a 3D modeling tool for crafting motion sculptures \cite{Kazi:2016:CFM:2858036.2858138}. Such tangible systems are frequently celebrated for being easy to learn, yet also having a high expressive ceiling~\cite{myers2000past}, which we hope to incorporate into our system.

Tangible systems have been particularly valuable in the context of animation. An early example of a tangible system for animation comes from the artists who worked on the classic film Jurassic Park. Their whimsically-named ``Dinosaur Input Device'' is described in a 1995 CHI paper~\cite{knep1995dinosaur}. The authors embedded sensors into an armature to control an on-screen puppet. The animators preferred the movement quality of the physically manipulated puppet as compared with computer interpolation-generated movement (and audiences did too - Jurassic Park is frequently cited as one of the best early examples of an animated character). Such tangible interfaces -- called ``maquettes'' -- are relatively common in the film industry; a recent example is the Baby Yoda character from Disney's The Mandalorian, controlled by four puppeteers with remote controls and one manipulating sticks connected to the arms. Glauser et al. created a system that allows animators to create tangible and modular rigs for controlling digital characters~\cite{glauser2016rig}, demonstrating improvements in accuracy and time with a posing task. Tangible puppets also enable capture for digital archive purposes~\cite{escobar2017tangible}. Dontcheva et al.~\cite{dontcheva2003layered} introduced a motion capture system that allows designers to use a variety of input methods as they rapidly prototype non-humanoid character motion. Gupta et al.~\cite{gupta2014motionmontage} enable animators to collage together multiple takes of an animation performance. Building on the benefits of tangible motion design systems, we draw from the field of puppetry to further augment tangible systems for animation design. Specifically, we seek to create a tool that focuses on scaffolding novice animators into rich, embodied explorations of movement, especially early in the design stage. \katherine{You do more than just "build on" these!!! I think you should be more explicit about the differences between your idea and these. I was particularly a little confused by how the Baby Yoda puppet was different. My impression is that your point is that it's important that the one animating is the one doing the hands-on exploration of form, which is something that the mentioned systems don't account for. I don't think that comes through very clearly}

\subsection{Sketching as a Design Practice}
Sketching -- whether with pen and paper or digital tools -- is a design practice that enables a creative practitioner to develop and refine ideas through an iterative process. Sketching -- especially early in the design process -- should support ``rapid, active and contextualized''~\cite{leiva2019enact} exploration and creation of a given design space. Creativity Support Tools (CSTs) that support early exploration can provide new \textit{paths of least resistance} for navigating a design space~\cite{myers2000past}. Fundamentally, the purpose of a CST is to support and extend the creative practitioner's relationship with her process, environment and tools. We embrace Dalsgaard's articulation of Deweyan philosophy, specifically the notion of \textit{instruments of inquiry}, an understanding of the way the creative process "intertwines" and "co-evolves with" the environment and tools. This elucidates the way a practitioner might leverage tools to augment her own cognition and creative process~\cite{Dalsgaard2014, PeterDalsgaard2017, Hollan2000}. These overarching concepts align with Sch\"{o}n's notion of reflection-in-action~\cite{schon1979reflective}. Specifically, our tool creates a \textit{path of least resistance} towards leveraging the physical world, and allows character designers to include diverse physical objects in their iterative brainstorming process.  

Hagbi et al.\cite{Hagbi2015ARsketching}
identify three `sketching' patterns: \textit{Sketching then playing} (where the sketch is a playing area for future gameplay), \textit{sketching as playing} (where the purpose of the activity is sketching - it is the main activity), and \textit{sketching while playing} (where participants alternate between sketching content and manipulating it). Our system embodies the ethos of `sketching while playing', but interprets the notion of 'sketching' more broadly, supporting 3D motion capture rather than on-paper drawing.

\subsection{3D Animation Tools and Techniques}
Most animation is carried out by following one of two classic animation techniques: key-framing (also called pose-to-pose) where the animator first defines specific poses along the animation trajectory, and then fills in the poses ``in-between'' these key poses using a process called ``in-betweening'' or ``tweening''. When animation is done digitally, this tweening may be done automatically by the computer, using a technique called \textit{interpolation} where the computer calculates the path between each pose and moves the relevant component along their respective paths. Interpolation has known issues with producing natural motion: namely the generated movements tend to be uncannily smooth~\cite{knep1995dinosaur}.

Another classic animation technique is known as ``straight-ahead''. This is the type of animation typically used in claymation or stop motion because it involves proceeding along the animation sequence linearly (rather than skipping ahead to future poses as happens in pose-to-pose). Some animators consider this more intuitive, especially for novice animators. A third technique is referred to as ``layered'' animation, and involves defining motion for collections of body parts separately, and then collaging the motion together in a final step. For example, K-Sketch allows the animator to create an animation of a wheel spinning while also moving forward along a path by allowing the animator to record both motions separately, and then automatically suggesting various combinations~\cite{davis2008kineticSketch}. Dontcheva et al. specifically designed a system around layered animation for motion capture~\cite{dontcheva2003layered}. Their system allows animators to perform different aspects of a moving character and layer these movements on top of each other to create the final animation. Our system similarly supports recording separate aspects of a digital character then combining them in a separate step. These techniques are complementary and are employed differently based on the animator's preference and the situation at hand.

As these traditional animation techniques were brought into computer graphics, designers began to create new systems, techniques, and tools for generating and capturing motion. One of the earliest examples of playing back an animation coupled to the motion of an input source was demonstrated in Baecker’s Genesys system~\cite{baecker1969interactive}. A later computationally mediated motion capture tool includes Calvert et al.'s Life Forms, the front-end of a more general-purpose 3D animation system that allows choreographers to use keyframes and inverse kinematics to create movement sequences \cite{Calvert-desktop-animation-1993}. Multi-touch has been shown to lower the barrier to manipulate complex characters~\cite{kipp2010multitouch}. Meador et al. used live motion capture to explore the role of mixed reality in a live dance production
\cite{Meador2004dance}. Procedural animation and physic simulation systems, including those built-in to Blender~\footnote{https://www.blender.org/} enable automatic generation of rigid-body, particle, and soft-body simulations. While the generated outcomes are often extremely compelling, there is less of a role for a human to design the movement in these procedurally generated animations. Most animation tools tend to support either key-framing~\cite{glauser2016rig, Calvert-desktop-animation-1993}, straight-ahead~\cite{knep1995dinosaur}, or layered~\cite{davis2008kineticSketch, dontcheva2003layered, ciccone2017authoring} animation techniques. While the final result created with any technique should be identical, the tools vary in how they support the ideation process. In this work, we seek to support an embodied, iterative, rapid prototyping process for animators creating character movement. \katherine{what about physics-based simulation/animation tools? Not super familiar w/ this lit, but I feel like a reviewer might think how tools like Blender (which you use) and its built-in rigid body simulation functions fits into this categorization} \rawn{yeah procedural animation seems like a thing to talk about, if only to say that it fails to support sketching in the same ways that these other tools do}

\molly{Authoring Motion Cycles~\cite{ciccone2017authoring}}.

\revised{
\subsection{2D Animation Tools} 
Motion artists often work with 2D interfaces, and researchers have developed myriad tools to support this process, \rawn{Researchers have developed myriad tools to support motion artists, especially novices, with 2D interfaces}aiming especially to support novice animators. Thorne et al. introduced a tool that allows animators to use a sketched gesture vocabulary to control 2D skeletons~\cite{thorne2004motion}.  Schiphorst et al. uses menus of postures to compose a movement sequence in time and space \cite{Schiphorst:1990:TIC:97243.97270}. \molly{Both?} systems are limited to pre-defined poses and 2D motion only while our system allows for more flexible, embodied authoring of 3D movement. \textit{Programming by demonstration} is a well-known technique for allowing designers to program application behaviour by performing it, recording it, analyzing it, and (in some cases) generalizing the behaviour across scenarios~\cite{lieberman2001your}. Animation tools regularly support this technique, which can be understood as a form of motion capture. For example, K-Sketch is a pen-based system that supports 2D animation-by-demonstration by novices~\cite{davis2008kineticSketch}. The authors of K-Sketch emphasized the importance of rapid, early ideation sketches~\cite{davis2008kineticSketch}. We resonate with their goals of creating a movement sketching interface that ``relies on users’ intuitive sense of space and time'', but introduce a tangible, free-form 3D based system rather than a pen-based system for 2D animation. \rawn{could also emphasize that yours is a system that works with the rest of the physical world - instead of being limited by what you can draw, you're limited by what kinds of things you can attach a controller to, which feels like a HUGE difference!}


A more performative animation tool is the video system from Barnes et al., that uses overhead tracking to capture the movement from a puppeteer manipulating objects on a desk and then digitally remove the hands to create a live, performative interface for cutout animation~\cite{Barnes2008VideoPuppetry}. Our tool is not meant for live performances, but instead supports the iterative design process, and focuses on 3D animation instead of 2D.  Willet et al. explored other ways to offload cognitive load from animation artists: their system automatically adds secondary motion to 2D animation \cite{Willett2017SecondaryMotion}. Another project involves editing and generating human motion by mapping glove-tracked hand motion to the entire body~\cite{lam2004motion}. 
These authors also seek to enable dynamic motion editing (rather than static keyframe editing). 
}

\revised{
\subsection{IGNORE FOR NOW}



An authoring tool for movies in the style of Heider and Simmel~\cite{gordon2014authoring}.

Miyoshi, K. 2019. ‘Puppetry as an alternative approach to designing kinesthetic move-ments’. In: Proceedings of the 4th Biennial Research Through Design Conference, 19-22 March 2019, Delft and Rotterdam, The Netherlands, Article 27, 1-16. DOI: 

"This paper explores movement and its capacity for meaning-making and eliciting affect in human-robot interaction. Bringing together creative robotics, dance and machine learning, our research project develops a novel relational approach that harnesses dancers’ movement expertise to design a non-anthropomorphic robot, its potential to move and capacity to learn. The project challenges the common assumption that robots need to appear human or animal-like to enable people to form connections with them. Our performative body-mapping (PBM) approach, in contrast, embraces the difference of machinic embodiment and places movement and its connectionmaking, knowledge-generating potential at the center of our social encounters. The paper discusses the first stage of the project, in which we collaborated with dancers to study how movement propels the becoming-body of a robot, and outlines our embodied approach to machine learning, grounded in the robot’s performative capacity." "HRI studies, however, consistently show that the more humanlike a robot appears, the more people expect it to also have humanlevel cognitive and social capabilities, so that interacting with these apparently humanlike machines often is frustrating and disappointing [4]. From a posthumanist viewpoint, this ambition to build mechanical servants and companions in our own image promotes not only humans making connections with machines but also eliminating human/machine difference [29]. It could be argued that robots mimicking humans or pets, often in cute, caricatured ways, deliberately blur the difference between organic and mechanical bodies, and human and machine cognition, to elicit human investment based on superficial and often false social cues. Designs that don’t rely on the familiarity of existing bodies, on the other hand, allow for human-machine encounters that are not restricted by "preconceptions, expectations or anthropomorphic projections ... before any interactions have occurred” [4]." ~\cite{gemeinboeck2017movement}

Relevant takeaway: Animated characters don't need to be humanoid or animaloid.
} 

\section{Design Motivation: Drawing from Expert Creative Practice}
\label{sec:expert-interviews}
As part of the design process, we engaged with professional creative practitioners in two related fields -- animation and puppeteering -- about their existing movement design process. We interviewed experts in both fields to develop an understanding of common approaches, techniques, and strategies for engaging in movement design. We identified both commonalities and differences, which allow us to identify fruitful opportunities for cross-pollinating the two fields.


\subsection{Interview Procedure and Analysis}
To gain an understanding of puppeteering and animation practice, we carried out semi-structured interviews with 2 expert creative practitioners. Interviewees were paid at the rate of $\$40$ an hour. The interview questions were guided by grounding themes of tool use, artifact generation, and personal creative practice, and shaped by the individuals' background and reflections.  Each interview lasted between 2 and 2.5 hours, during which we asked a semi-structured set of interview questions, focusing on their personal creative practice and background. Questions included probes about the creative process such as \textit{"Can you walk me through your design process for a particular project?"} or \textit{"What role does this technique/strategy/approach play in your creative exploration?"} or \textit{"What are the benefits of technique A in comparison with technique B?"} We then performed thematic analysis on the interview transcripts, iteratively reviewed and analyzed all interview data and discussed all emerging themes~\cite{mcdonald2019reliability}. Themes are presented below, categorized into strategies these practitioners use to structure their respective creative processes.

\subsection{Participants}
Both participants were recruited via professional contacts, and invited to speak with the lead researcher while remotely connected over a video-conferencing system.  

\textbf{\textit{Animator}} --
The animator has been working professionally as an animator for over 11 years. She now works for a large animation studio, and has worked with many different companies throughout her career. She works primarily in 3D, and has also explored 2D, stop-motion, and VR animation.

\textbf{\textit{Puppeteer}} --
The puppeteer is an Emmy award-winning performer who has been performing professionally for over 25 years. His work could be categorized as physical comedy, clowning, mime - he excels at physical performance. His primary puppeteering technique is Found Object puppeteering, where everyday materials are manipulated with the hands to create compelling characters. For example, a plastic bag may fold and slightly inflate to become the body of a chicken, with plastic forks for feet.

\subsection{Findings}
Experts in their respective fields, both our informants have rich practices of movement design. We identified both similarities and differences in their techniques, and identify opportunities for importing expertise from puppeteering into animation, which could shape the design of tools to help scaffold newcomers.  

\subsubsection{\textbf{The importance of texture and rhythm}}
Both the Animator and the Puppeteer are highly attuned to movement qualities such as rhythm and texture as they design characters, develop scenes, and tell stories through their respective mediums. 

\squote{Animator}{I don't make everything smooth in the scene. I try to include staccato movement to give more rhythm... 
I use some motions more straight and then some motions more like round shapes.  
}

The Animator was highly attuned to movement qualities like rhythm (e.g., staccatto, smooth) and texture (e.g., sharp, round) and how these would shape the final outcome. Similarly, the puppeteer heavily emphasized Laban movement concepts, specifically the four categories \textit{sustained}, \textit{pendular}, \textit{abrupt}, and \textit{vibratory}\molly{CITE}. As he's developing a character or a scene, he keeps these terms top of mind, using them to shape his rehearsal process, and iterate on the design of a character.

\squote{Puppeteer}{
The real kicker is the transition between 
multiple states. So you create a low vibration going into 
a high pendulum movement. 
It's surprising to see that shift of the two different energy levels and that's what people respond to.}

Both experts emphasized the importance of texture and rhythm throughout their design process, highlighting attention to detailed aspects of movement quality as a shared value.

While both experts highly valued nuanced movement qualities, they had different relationships to the process of generating such movement. The puppeteer's Found Object puppetry technique emphasizes the use of physical materials to create characters. A paper bag can be expanded with the hands, then crumpled, then re-inflated to convey breath. A piece of foam or a scrap of fabric might be stretched out to communicate gravity and weight. The puppeteer articulated the ways in which these materials shape his exploration of characters:

\squote{Puppeteer}{[The character design] all depends on the material that you're using...you start playing with it and then that's where you make the discoveries.}


The materiality inherent in his puppeteering practice guides him throughout his design process and helps him generate new movement qualities. In contrast, the Animator's digital process has a more limited relationship to materiality.
She discussed the particular challenge she would face if asked to design an abstract, non-humanoid, non-animal character:

\squote{Animator}{If I had a character that was the shape of water 
I would just try 
to [create] motion in general, I guess, rather than looking for a reference... thank God I have never had to do that -- I probably would have a hard time.}

In contrast, the puppeteer 
was able to rely on exploration with tangible materials to aid his ideation as he generates motion ideas for such abstract characters. In general, the physical objects played a major role in the Puppeteer's design process, the materials directly influencing movement:

\squote{Puppeteer}{
[The way we move] becomes unconscious, becomes habit, becomes muscle memory, becomes us. 
[Using a puppet provides] 
a sense of allowing your body a chance -- and your mind a chance -- to shift its perspective. 
You add new limitations onto yourself and create new avenues for yourself.
}

He articulated the value of a physical object: it provides additional movement constraints, suggests new movement qualities because of those constraints, and provides some ``separation from oneself'' throughout the creative process. While tangible ``puppets'' -- called maquettes -- were frequently used in the early days of animation~\cite{knep1995dinosaur}, they are less frequently used now, and the Animator had not used them in her 3D animation projects. Her design process centers around digital characters, which have no analogous qualities.

\subsubsection{\textbf{Observation and attention to detail}}
The Animator also described the ``misconception'' that animators often feel early in their career, when they incorrectly believe they can simply generate natural movement without first engaging in focused observation: 

\squote{Animator}{
Even right now, you talking to me -- you think you know what you're doing, but your shoulder is moving and your head is nodding and you don't notice the frequency at which it's nodding.
We think we do, but we don't. 
You need help to go to the source of the movement and seek the truth of the animation, which is recreating life.}

Both practitioners articulated strong attention to the way subtle and nuanced movement design influences the final character. \molly{maybe need a different conclusion here- one that ties observation to physically generating movement in the moment, during the sketching phase.}


\subsubsection{\textit{\textbf{Summary}}} -- 
In summary, movement qualities are very important to character design in both contexts, and the Puppeteer finds that incorporating physical materials into his character design process is an extremely effective method for positively influencing the movements and characters he generates, especially for abstract characters. As we reflected on these interviews, we generated the following questions, which this paper seeks to address: \textit{Would an animator get similar benefits from incorporating materials into their design process? In what ways might the Found Object puppeteering technique positively influence an animator's style, or exploration process? How might we design a computational system that allows such material explorations? Would a new or augmented animation tool fit into an animator's existing workflow?} In this work, we seek to answer how we might design a computational system which incorporates the material exploration that the Puppeteer found so essential.

We also discussed the role of computational tools in animation with the Animator, who repeatedly emphasized the importance of developing skills in the ``art form of animation'', and of not getting hung up on the specific animation software in use. She discussed the way novice animators are sometimes undermined as they begin their journey to learn animation:

\squote{Animator}{
We all say “animation is recreating life” but then the first thing that new animators do is get in front of a computer and try to learn software -- they forget that life aspect. 
}

We interpret this as a call to the importance of observation of the physical world in animation. The combination of the Puppeteer's physical materials -- or jigs, because they helpfully constrain movement -- and the Animator's desire for novice animators to develop an eye for movement suggests the potential value in incorporating physical objects into the animation process. 

\katherine{I unfortunately don't have a super concrete suggestion, but I'm not sure I'm sold on this section. I think it's a valuable addition, but somehow the breaking up of your findings into 4 themes seemingly on equal footing is a little confusing for me in the context of this paper. I believe the bottom line, as you say in the summary, is that there are shared values between both practitioners (around rhythm, texture, and the attention to subtle features that make movement natural), but the puppeteer's unique focus on materiality guides him throughout his design process and helps him generate new movement...and you wonder what it might bring to the digital animator's life. This doesn't come through to me in the current structure. I think at the very least, reordering the themes would help -- maybe present Rhythm/Texture and then "Recreating Life" first (the title of the latter is also slightly confusing -- I'd hate to use the word "Importance" too many times, but this section feels more like speaking to the "importance of observation/attention to detail" or something)?} 

\sarah{I agree completely.  A couple things to try - get ride of the 3.1 and 3.2 header, and make that one "methods" subsection, since they're so short.  

In the findings subsection, what if you told the stories as Animator, then Puppeteer, then drew the thematic similarities at the end?  As is, it feels really disjoint, I'm not clear on why these two interviews were combined in the first place, or why in some sections you only talk about 1 person.  But if you were like "here's somethings an animator cares about," then "here's how a puppeteer approaches movement," then "look at these similarities and differences", it might flow better?}

\rawn{I think the summary paragraph could use more work- I think the questions you asked are really excellent - can you frame them as "from these interviews, we came away with questions like" or "we endeavored to answer how we might design a computational system which incorporates the material exploration which the pupetter found so essential and the animator seemed to be striving after"}



\section{\systemTwo}
Throughout the formative interviews, we were struck by the role that materials played in the Puppeteer's process. While both experts discussed the importance of varied movement qualities, we wondered how much the digital ``material'' of 3D animation influenced the final outcome for those using 3D animation tools. The core idea behind \systemTwosp is finding a way to incorporate the same types of physical materials used by the Puppeteer into digital animation workflows in order to enhance sensitivity and attention to the physical movement qualities valued by the Animator.

The Puppeteer described his materials as helpfully constraining his movements, much like the way that jigs and fixtures support woodworkers by providing selective constraints to motion. While jigs in woodworking are typically solid and hold cutting materials securely in place, and the Puppeteer used soft or flexible materials such as a napkin or a jacket, the core idea of an external tool that helpfully limits movement remains the same. We therefore also refer to the materials we incorporate into the animation process as \textit{jigs}. \rawn{is this the same paragraph from the intro?}

In order to incorporate physical jigs into the animation process, we needed to develop an animation system that  1) provides extremely precise controls for manipulating digital characters, and 2) allows for the incorporation of physical materials that may occlude the hands into the animator design workflow. To fit into existing animation workflows and maintain ecological validity, our tool should utilize familiar rigging, modeling, and animating software. To satisfy this last requirement, we use the open-source 3D CAD application Blender for the rigging and modeling of our digital characters. The second requirement eliminates many otherwise promising solutions: even state of the art hand-tracking libraries are not designed to work when the hands are occluded, for example. Instead, we use HTC Vive controllers, which are accurate even when partially or mostly occluded, and provide a more robust connection between the digital and physical worlds. The HTC Vive controllers also provide centimetre-level accurate tracking, thereby satisfying the first requirement especially for an early sketching tool~\cite{buxton2010sketching}. By displaying the model on a desktop monitor we eliminate any need to repeatedly remove and replace the headset (a frustrating barrier to creation in the world of VR~\cite{thoravi2019tutorivr}) which would violate our original goal of creating a rapid sketching tool. Below we describe the technical architecture of \systemTwo, and the evaluation we carried out to assess the system.

\begin{figure}[t!]
    \centering
    \includegraphics[width=.92\columnwidth]{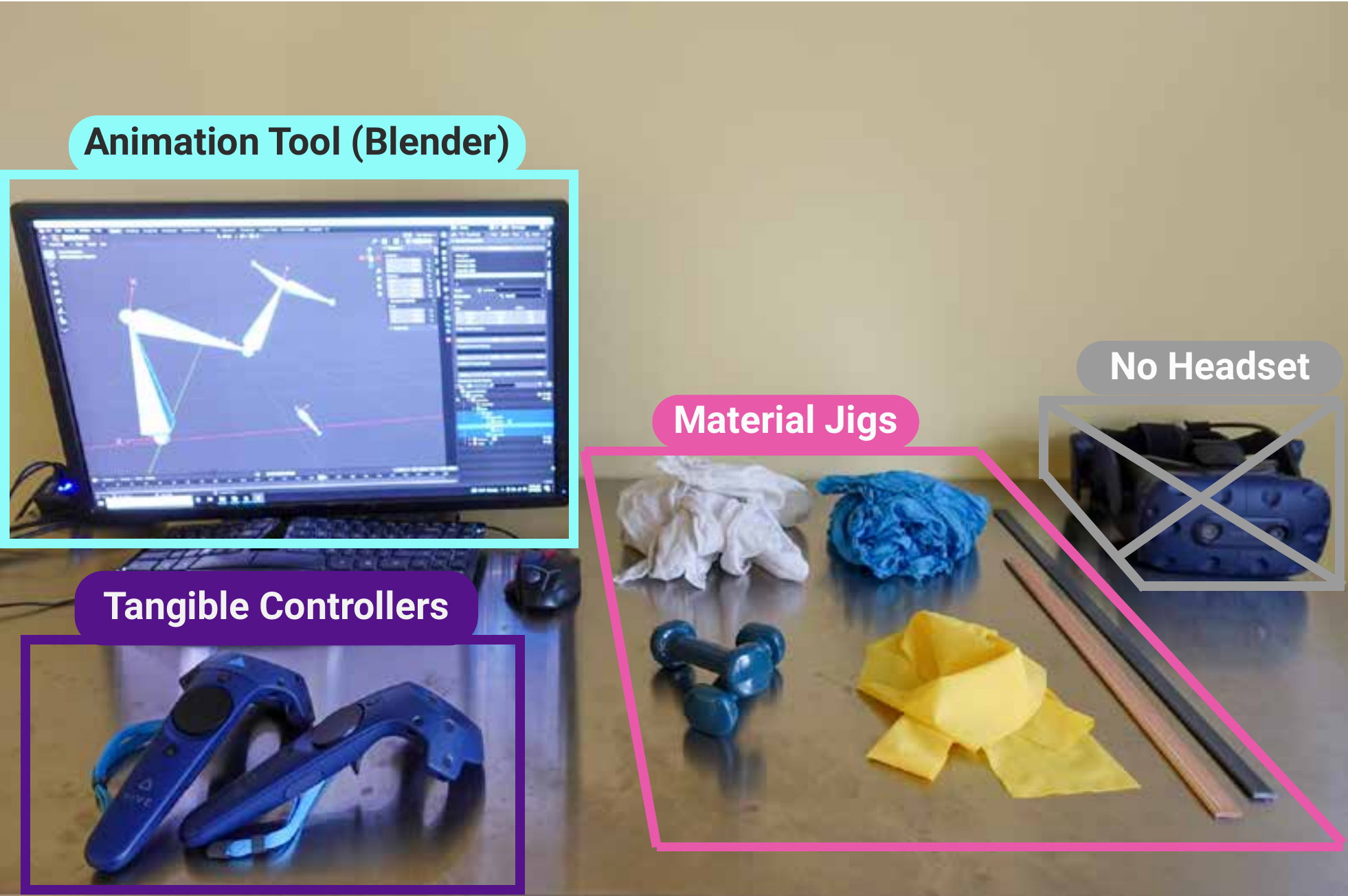}
    \caption{\systemTwosp consists of 1) the animation tool Blender running on a desktop computer, 2) HTC Vive controllers, allowing tangible manipulation of the digital model, and 3) material ``jigs'' for the designer to manipulate during ideation. Note that no headset is required: the digital puppet is viewable through the desktop computer monitor.}
    \label{fig:system}
\end{figure}

\subsection{\systemTwosp Technical Architecture}
\systemTwosp is a system that enables animators to manipulate a digital character while also interacting with physical material jigs, or otherwise taking advantage of features in the physical world (e.g., gravity, momentum). It shares important features with other motion capture systems: both our system and other motion capture systems allow animators to capture performed body movement. However, there are important distinctions: our system does not require a full-body tracking outfit, instead the only tracked components are the controllers. This makes tracked movement simpler to generate and allows for quicker transitions between performing and editing, key to any creative process~\cite{simon1969sciences}. 
Our system is designed to support rapid, early ideation in the pre-production character design process where designs are meant to stimulate conversation; any generated character movement is not meant become part of the final production. In this way, the generated movement can be understood as an early sketch, and is not meant to be highly polished or complete.

The system consists of three components (see Figure \ref{fig:system}): 

\begin{itemize}
    \item The Animation Tool -- this provides access to the rig and 3D model. The Animation Tool is Blender, an open source 2D/3D content creation tool~\footnote{Version 2.93, \url{https://www.blender.org/}}, running on a desktop computer. The model can be adjusted with either the keyboard and mouse, or the tangible controllers. Blender also provides basic animation functionality such as keyframe recording, rigging controls, inverse kinematics, etc.
    \item The Tangible Controllers -- Two HTC Vive controllers provide centimetre-level accurate position tracking, and act as a translator between the physical world and the digital one. Using a custom script\footnote{\textbf{The URL of the open-source software will be provided after acceptance.}} and the opensource library PyOpenVR\footnote{\url{https://github.com/cmbruns/pyopenvr}}, we stream location data from the controllers into Blender, where they change the location and orientation of a selected character.
    \item Material Jigs -- based on the Puppeteer's described technique, we collected a variety of physical objects for animators to use as jigs during their animation process, including weights, a stretchy band, different kinds of fabric, and a plastic bar (see Figure ~\ref{fig:system}).
\end{itemize}

To use the system, an animator first creates or downloads a 3D rig (a standard first step in all animation). Next, the animator opens the custom UI \molly{add figure}, and connects any single bone in the armature to each controller. Now the controller movement is bound to the 3D rig, which allows the animator to control the digital character with physical movements in the real world. Optionally, the animator can use the material jigs to influence, perturb, inspire, and shape their movements, similar to the way the Puppeteer used materials in his design process (see Section \ref{sec:expert-interviews}).

\begin{figure}[t]
    \centering
    \includegraphics[width=.82\columnwidth]{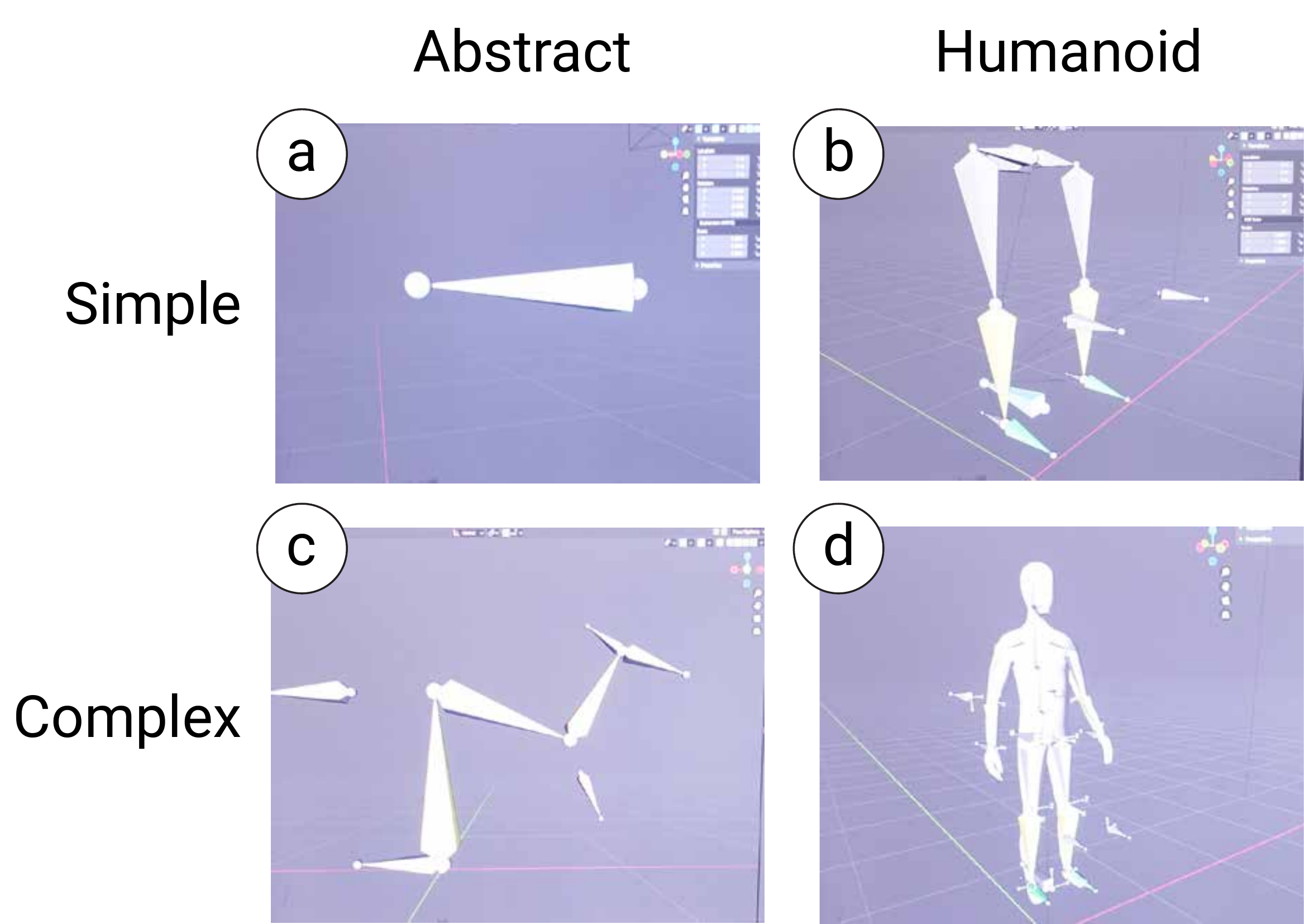}
    \caption{Four 3D models provided to participants during the user study. a) the simple abstract model provides almost a direct mapping to the controller, providing a good baseline for understanding the system. As expected, participants immediately ascribed personality and a story to even this simple geometric shape~\cite{heider1944experimental}. b) Two ankle bones and a pole target located at the knee control these humanoid legs. c) the complex abstract character uses inverse kinematics to move two bones by controlling a bone connected to the head bone. d) The full humanoid includes more than 5 control bones, and multiple pole targets providing motion constraints. \katherine{You don't ever reference b) or d), so their inclusion here is somewhat confusing}}
    \label{fig:model-options}
\end{figure}

\section{\systemTwosp Evaluation}
Evaluating novel toolkits is notoriously difficult~\cite{olsen2007evaluating}, sometimes -- as in the case of usability assessments -- even considered harmful~\cite{greenberg2008usability}. Beyond usability evaluations, there are a variety of strategies that can be used to assess toolkit effectiveness~\cite{ledo2018evaluation}. Like many novel toolkits, \systemTwosp requires time and effort to build familiarity and incorporate into a workflow, meaning such a tool would not typically be considered a good candidate for lab usability studies~\cite{ledo2019astral, greenberg2008usability}. Instead, the focus of our user study was not on usability, but rather on understanding how incorporating physical elements as a first-class design material alters the design decisions taken by practitioners even during a first encounter. We therefore invited two novice animators in to experience the tool. \rawn{so the initial framing was making animation software easier to learn (at least that was how I read it), but the evaluation is about how novice users make more interesting design decisions with physical materials (a much more fascinating research goal imo!), so maybe tie that back to the beginning}

\subsection{\textbf{Procedure}}
Both participants visited our lab for a 1.5 hour workshop and was compensated $\$40$. Each session consisted of 1) interview on background and personal design practice, 2) a warm-up tutorial 3) a series of exploratory design tasks following a think-out-loud protocol and 4) a post-study interview. Participants were introduced to the tangible controllers first, while learning to control a single bone (see the simple abstract model in Figure \ref{fig:model-options}a), and learning how to move the controller in physical space to control the digital character. Next, they were introduced to the notion of the physical jigs: \textit{``When puppeteers design a new character, they often play with different materials as they're exploring. We've provided these different materials for you to use as you think about the character you are designing. Interacting with a physical material might influence how the final movement looks''}. After participants had experienced the simple abstract character and the jigs, participants proceeded to the exploratory design task. Participants were instructed to iterate on a new character design, and to create two 5-10 second scenes (one with low energy and one with high energy) where an audience would learn about that character through the way that they move. Both participants chose to design their movement using the complex abstract character (see Figure \ref{fig:model-options}c). During the study, participants chose which found object material jig(s) to use while holding the controllers and manipulating the on-screen digital character.

We recorded and transcribed what each participant said while thinking-aloud as they experimented with the tool for the first time. We then performed a thematic analysis on their transcribed quotes, and synthesized our findings into themes. This study design allowed us to observe the way the tool affects the design process on a first encounter, with designers who are new to the system.

\subsubsection{\textbf{Participants}} The study was conducted with two~\footnote{Due to a spike in COVID-19 cases in our area, we unfortunately had to cancel the vast majority of scheduled participants.} novice designers (avg. 29 years of age, 2 female). Participants were recruited from university mailing lists in Art, Architecture, Design, and Computer Science. Prior experience with 3D modeling was self-reported in a preliminary survey; we purposefully recruited participants with varying levels of expertise in animation: one participant reported intermediate experience with animation, and the other participant had no prior experience. Since we have a small number of participants, we describe them in more detail to further contextualize their responses:

\textit{P1} - P1 is learning Blender, and has intermediate experience - she has used it for several ongoing animation projects. Her background is in product design, and while she has primarily worked on websites up until now, she is very interested in tangible experience design.

\textit{P2} - P2 has a background in psychology, and has zero prior experience with animation, or animation tools. She is a film buff, and considers herself well-versed in animated movies as an artform, but has never created animation of her own.

\subsection{\systemTwosp Study Results}
Even in a brief workshop-style experience with limited exposure to the tool and this method of working with digital animation, users readily engaged in unique movement-generating behaviour. See Figure \ref{fig:teaser}, right for examples of jig exploration our participants engaged in.

\subsubsection{\textbf{Access to Jigs Influenced Design Process and Outcome}} P1, who has some prior experience designing animation in Blender, compared her experience using \systemTwosp with Blender. In particular, when designing a ``low energy'' experience for her character, she experimented with weights as jigs. P1 connected the feeling of heaviness with the increased weight of the emotional message she was hoping to convey:

\squote{Designer}{
Low energy 
means I have a lot 
buried in my shoulder and in my mind. 
That's how it 
feels - your body is very heavy. I just want to see 
what that would do to my hand and my character if there's actually 
weight on it.
}

While she was familiar with Blender's built-in parameter to increase the weight of a character's body part, the experience of physically manipulating weights as she performed the digital puppet's movements impacted her design experience. She described the way she might update the ``weight'' of an object in Blender, and compared that with the experience of holding varying amounts of weights while animating with \systemTwo, which she described as ``the real version'' of such a design choice. The tool created an embodied experience with weight:


\squote{Designer}{
[This tool] is a way to embody 
when I change the metrics or parameters in the software. 
Without this tool, it was just a click 
from the mouse and 
it doesn't feel that real.  
I thought it was real before, but now, with this, I feel like this is 
great – way more real!}

P2 also engaged in extremely physical exploration of the tool, moving around the room, waving her arms in the air, bouncing the controller on the fabric, and swinging the controllers around. She felt that the material jigs anthropomorphized the movement that the tool helped her create:

\squote{Novice}{
Using this free-flowing and bouncy material 
almost personifies this figure in a way. 
If I use it just with my hand it’s more controlled 
but this makes [the movement] more unpredictable 
}

P2 found this controlled unpredictability an appealing and compelling addition to her character during the creation process.

\subsubsection{\textbf{Tangible Control System Enabled Physical Explorations}}
In addition to using the jigs to modify her character's movement qualities, P1 experimented with using the tangible controllers to incorporate gravity and momentum into the digital character movement. She tied the controller to a piece of stretchy fabric, and let it drop as an expression of despair:

\squote{Designer}{
I like how it just hangs here. Because when you drop everything you're like “I don't have any hope - I’m so sad, no, no.” 
}

Similarly, P2 described a compelling sense of less control when using the material jigs, which she felt improved the movement:

\squote{Novice}{I really like using these [materials] 
because you have more range and it 
comes to life 
more.
It's just less structured.}

Both participants felt the material jigs positively affected their overall design experience.


\revised{
\section{\systemOne}

\begin{figure}[t!]
    \centering
    \includegraphics[width=.8\linewidth]{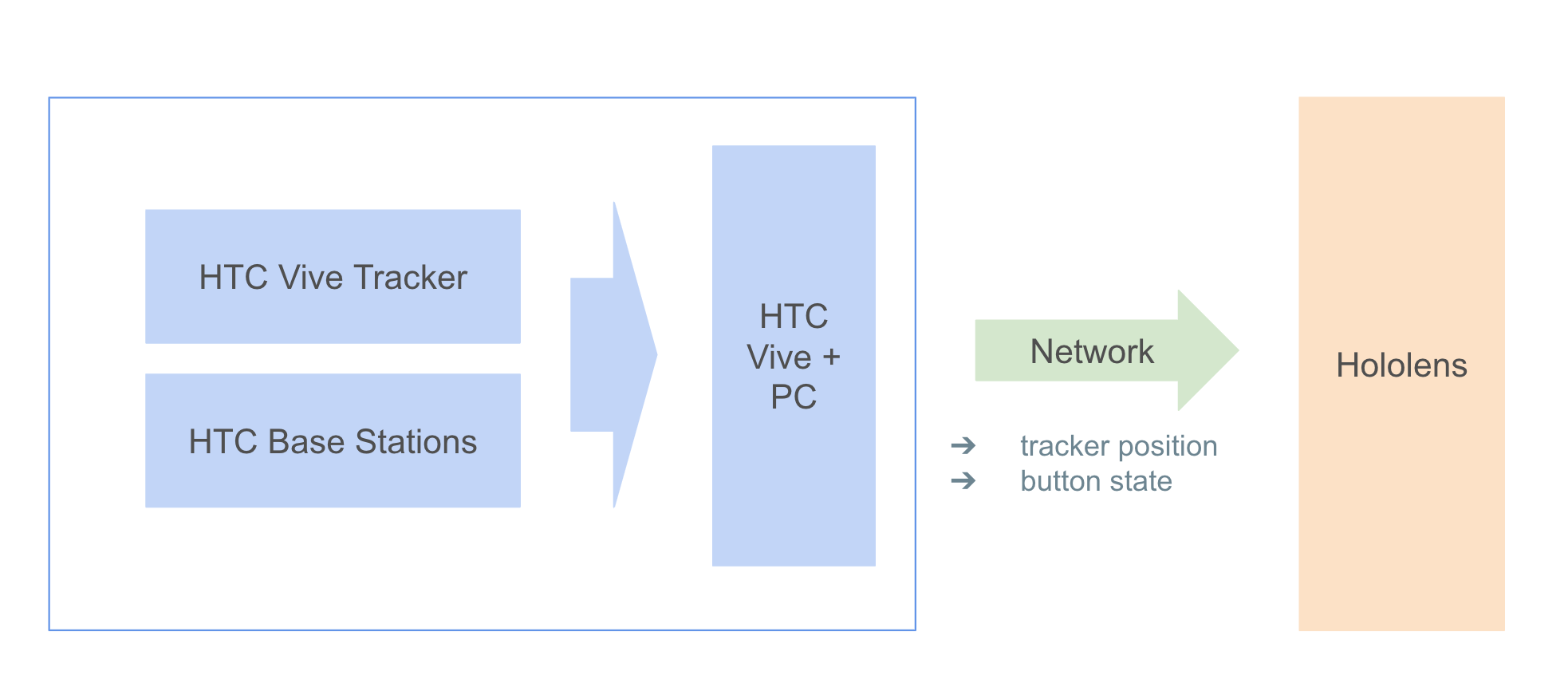}
    \caption{HTC Vive sends tracker position and button states to Hololens via network.}
    \label{fig:system_communicationS}
\end{figure}

\molly{needs more philosophy, goals, astral-style motivation. Also need to motivate from RW, so we can motivate the interviews w/ expert practitioners.}

As part of our evolving design, we first built a system that leverages tangible controls, and a head-worn display, called \systemOne. 

A choreographer designing for a stage typically needs to be able to view the space at a variety of angles and distances. Kim et al. built a scaled down version of a studio to aid in smart home prototyping \cite{Kim:2016:MDT:2858036.2858180}. Similarly, our tool supports `zooming out' from the collection of trajectories to simulate viewing from a more distant vantage point in addition to translation and rotation.

A thorough overview of computational tools used in choreographic motion planning identifies four `types' of technological systems, organized by purpose: \textit{reflection}, \textit{generation}, \textit{real-time interaction}, and \textit{annotation} \cite{Alaoui2014Choreography}. We position our tools as `generative' tool since they are designed to be used during the `development' stage of the creative process \cite{Singh2011ChoreographerNotebook}. The term "generative" can refer to autonomous algorithms that computationally produce movement trajectories within set parameters, but we are referring to the more general, high-level definition common in creative practice of creating or 'generating' work. Like the Dynamic Brush system from Jacobs et al., our tool is designed to extend the manual skills of a movement creator as they \textit{generate} movement rather than replace those skills \cite{Jacobs2018brushes}.

Singh et al. created a multi-modal annotation tool that allows dancers and choreographers to collaborate outside of the normal rehearsal space \cite{Singh2011ChoreographerNotebook}. Our work focuses on the authoring step, but could be extended to allow remote collaboration and annotation. The majority of choreography-support tools focus on annotation rather than creation.

\subsection{\systemOnesp Technical Architecture}
\systemOnesp is comprised of two parts: an HTC Vive and a Hololens.
Figure \ref{fig:system_communicationS} explains the data flow.

We use the HTC Vive for hand tracking and user input.
Users hold two HTC controllers that are automatically tracked by the HTC Lighthouses for cm-accurate precision location tracking.
The HTC headset is required to read data from HTC trackers; furthermore, it handles the network connection coming from Hololens, but is otherwise unused.

The Hololens displays virtual objects in AR, and acts as the Heads-Up Display (HMD) for the system.
The Hololens connects to the HTC Vive through a local area network, reads tracking data from the HTC Vive, and displays virtual objects.
Hololens also handles the computation (rotation, translate, scale) and state management.

\subsubsection{Tracking} -- Accurate tracking is fundamental to our system, which requires a tracking system that is accurate while providing enough freedom for users to perform and be expressive.
Hololens's hand tracking system~\footnote{https://docs.microsoft.com/en-us/windows/mixed-reality/gestures} works only when the hand is visible to the device, which would undermine the performative power of such a movement tracking system. Instead, we chose the HTC Vive for hand tracking because this solution provides excellent accuracy~\cite{niehorster2017accuracy} and sufficient room-scale freedom of movement. 

\subsubsection{Calibration} -- The \systemOnesp system requires a calibration step to translate HTC Vive's coordinates into Hololens's coordinates.
We assume that, given a coordinate $x\in\mathbb{R}^3$ from HTC Vive's coordinate system, the corresponding $x'$ in Hololen's coordinate system is
$$x' = kAx + b$$
where $k\in R$ is a scaling factor, $A\in\mathbb{M}_{3\times 3}$ is a rotation matrix, and $b\in\mathbb{R}^3$ is a translation vector.

Our simple calibration procedure is described below. The variable $t$ represents a reasonable unit of distance that is up to the choice of the system designer.
\begin{enumerate}
    \item User aligns the HTC tracker with a cube at $(0,0,0)^T$ in HoloLens space, which gives a reading of tracker's position $x_0$ in the Vive space.
    \item User aligns the HTC tracker with a cube at $(t,0,0)^T$ in HoloLens space, which gives a reading of tracker's position $x_1$ in the Vive space.
    \item User aligns the HTC tracker with a cube at $(0,t,0)^T$ in HoloLens space, which gives a reading of tracker's position $x_2$ in the Vive space.
    \item User aligns the HTC tracker with a cube at $(0,0,t)^T$ in HoloLens space, which gives a reading of tracker's position $x_3$ in the Vive space.
\end{enumerate}
Then, let
$$a_1 := x_1 - x_0$$
$$a_2 := x_2 - x_0$$
$$a_3 := x_3 - x_0$$
Given a reading of tracker's position $x$, the Hololens's position is given by
$$x' = t\left( \frac{(x-x_0)\cdot a_1}{\|a_1\|^2}, \frac{(x-x_0)\cdot a_2}{\|a_2\|^2}, \frac{(x-x_0)\cdot a_3}{\|a_3\|^2}\right)^T$$

In this algorithm, we choose three orthogonal vectors as the basis of the HoloLens's coordinate system.
A reading from HTC Vive's coordinate system is projected onto each direction, translated into Hololens's coordinate system, and recomposed as the position in HoloLens's world.
We choose $t=0.1$ to keep the calibration cubes close to each other.

\subsection{\systemOnesp Interaction Design}
When designing the user interaction experience for \systemOnesp, we placed a high priority on embodied interaction, direct manipulation, untethered movement and a closed loop editing tool. One of the more frustrating experiences when working with VR/AR applications is having to frequently take the headset on and off to test. In VR, this frustration is compounded by the fact that the real world is not typically visible to users. 

Our motion path creation tool \systemOnesp lets users move freely within the real world while also allowing for embodied interaction with the virtual environment and direct manipulation of their creations.  This creates a rapid feedback cycle that is critical to our goal of providing a low-fidelity sketching tool for generating motion paths.

\subsubsection{\textbf{Creation}} --
An advantage of relying on HTC Vive trackers for drawing trajectories is that users can use the entire space around them to perform. In contrast, with HoloLens tracking alone, users would have to hold a tracker directly in front of their faces while they created.

To start a trajectory, users tap the right trigger and then move their right controller anywhere within the tracking area. While they are drawing their trajectory, the program keeps track of timing data so that slowly drawn trajectories generate waypoints in close proximity, and quickly drawn trajectories will have waypoints that are spread apart. To stop recording, users simply tap the right trigger again. From here, users have the opportunity to move around within the space to see their trajectory from different angles. Then, users can choose to create another trajectory, edit trajectories, or replay trajectories.

\subsubsection{\textbf{Editing}} -- After creating a trajectory, users can edit the trajectory in three ways. These editing modes are accessed through a menu where users select the trajectory or trajectories, the editing mode, and the axis (if rotating). We created this menu to replicate some aspects of current animation tools like Blender in an attempt to allow for a more direct comparison against those tools. This design decision is further examined in the Discussion section.
\begin{itemize}
    \item Translate: Users should feel as though they are manipulating the trajectory by grabbing it. Users touch the right touchpad and then move the controller to translate trajectories in three dimensions proportional to the movement of their controller.
    \item Rotate: Users tap left and right on the left touchpad to rotate trajectories along their chosen axis. X is "right", Y is "up", and Z is "out". The left and right side of the touchpad are used to control clockwise or counterclockwise rotation.
    \item Zoom: Users tap left and right on the left touchpad to zoom trajectories in and out. The zoom function works as an expand/contract function, making points either further away or closer to the center of the trajectory.
\end{itemize}

\subsubsection{\textbf{Replay}} --
The replay function allows users to see their generated movement replayed. Replay sends a cube along the trajectories at a speed proportional to the speed at which the user drew the trajectories. As the cube moves, all other waypoints become transparent except for the five waypoints in front of the cube. This allows the user to track movement even in a space busy with large or complicated trajectories. 

\subsubsection{\textbf{Layered trajectories}} -- Users can stop the replay at any point and add other trajectories. This allows for the creation of layered movements that may start and end at different times, creating an overlapping pattern of movement.

\subsection{\systemOnesp Evaluation}
We invited 5 participants to participate in a within-subjects study where they compared our system to the popular 3D modeling and animation tool Blender, and provided feedback on the experience of using both tools. The goal of our user study was to understand how enabling the use of the body in a digital environment alters the design decisions made by practitioners.  We first surveyed our participants for their current design practices, then had them carry out the same task on both our system, and Blender. This study design allowed us to describe the material conversations that occurred with the system, as well as some initial system usability.

\subsubsection{Participants}
Participants were recruited from the university campus Design mailing list.  One participant did not show up, so we recruited an additional participant from the Makerspace where the study was taking place. Participants were selected to represent a diversity of creative practices and experience levels to maximize the range and diversity of experiences:
\begin{itemize}
    \item P1: Physical Animator: one year of experience doing stop-motion animation with clay, no other movement experience, 3 years doing model design with Fusion 360.
    \item P2: Choreographer: more than 2 years of experience with performing and choreographing traditional forms of Chinese dance, minimal experience with 3D animation tools.
    \item P3: Dancer and 3D designer: 5 years of dance experience and 5 years of experience using 3D design software, including Maya.
    \item P4: AR interface designer: experience designing authoring systems in AR, and working with drone trajectory paths.
    \item P5: Roboticist: 7 years of experience controlling drone swarms and a jumping robot.
\end{itemize}

\subsubsection{Procedure}
Participants were invited to meet with us in our studio space for an hour-long workshop and paid \$20/hr. All participants self-reported zero prior experience with both Blender. Each session consisted of 1) a survey participants on personal design practices (in both physical and digital contexts), 2) a warm-up tutorial 3) create two motion paths in Blender, 4) create two motion paths with our system, 5) survey with a subset of the Creativity Support Index~\cite{Cherry:2014:QCS:2633907.2617588} (we omitted questions on collaboration since they were not relevant the current study). Throughout the study, participants were encouraged to follow a ``think-aloud'' protocol. 

Participants were asked to imagine they were designing a drone performance for a stage show. In both tools, they were instructed to create the path two drones would take from a landed position on a piece of furniture (in Blender a mesh cube and in our system a stool) placed in the center of the ``stage''. 

\subsection{\systemOnesp Study Results}
We first present qualitative results, collected from the recorded think-aloud utterances. We carried out a thematic analysis of responses, which we synthesize into common themes and insights to support the
design of future tangible interfaces for digital manipulation. Then, we share quantitative results with descriptive statistics due to low sample size.

\textit{\textbf{Embodied interface affords complexity}} --
In both conditions, participants expressed a desire to generate complex paths with many turns and curves to make the paths ``more interesting''. In Blender, participants quickly became discouraged:

\squote{Choreographer}{This is way harder than I thought.}   
\squote{Dancer}{I'm not sure [my design is] possible with these tools.} 
\squote{AR designer}{Clearly I couldn't get my design here.}

While using Blender, participants tended to struggle with creating 3D shapes in the 2D environment. All participants would frequently rotate their view around while attempting to translate their trajectories, often verbally remarking on the difficulty of ``trying to move something in 3D while looking at it in 2D''.

In contrast, while using \systemOnesp participants simply moved their bodies physically around the space to see how their trajectory was placed:

\squote{Choreographer}{You can just do it with your hand instead of trying to do it in 2D space with a computer.}

With \systemOne, all participants indicated they were better able to create a design with which they were satisfied: 
\squote{Roboticist}{You get a trajectory right off the bat that is what you want.}
\squote{AR designer}{I was trying to draw hearts in 3D...the authoring part was straightforward}

Two participants using \systemOnesp indicated that they had no need to edit and were satisfied with the trajectories they had created.

Participants tended to create much simpler trajectories in Blender, and much more complicated designs in our tool, even though both tools were new to all participants -- further demonstrating the way our embodied interface supports immediate engagement.

\textit{\textbf{Low threshold, unique strategies}} -- Physical skill or experience moving through space did not affect satisfaction. Instead, our tool was immediately accessible to all participants (low \textit{threshold}~\cite{myers2000past}), regardless of movement background \molly{add threshold and ceiling discussion to intro / rw}. Participants shared different strategies for generating movement. The roboticist, with only minimal movement experience, wanted to create sub-sequences of a trajectory and chain them together rather than creating them all in one gesture. In contrast, the choreographer preferred to completely perform her movements in full, and wanted to see her trajectories in high fidelity. The other participants, with varying amounts of movement experience, requested more editing functionality such as the ability to `cut' a trajectory and bounding box style scaling.


\textit{\textbf{Quantitative Results}}
Our participants filled out a standard CSI questionnaire, answering questions related to Enjoyment, Expressiveness, Results Worth Effort, and Exploration~\cite{Cherry:2014:QCS:2633907.2617588}. The questions were on a 10-point Likert scale, with 1 being Strongly Disagree and 10 being Strongly Agree. \systemOnesp scored highest on Expressiveness and Exploration scores, and lower on Results Worth Effort and Enjoyment. One participant elaborated on the discomfort of wearing the Hololens headset, which caused them to rate the system lower on Enjoyment. The survey results matched our observations: participants felt that the system did a better job of supporting exploratory behaviors than the desktop system. See the Discussion section for more details.

} 

\section{Discussion}
By incorporating the material jigs into the design process, our participants developed their sense awareness~\cite{ghefaili2003cognitive} of the physical world. That is, in addition to choosing jigs to influence, shape, refine, or limit their movement, participants also began to experiment with the way other physical elements such as gravity and momentum could influence their character's motion design. This increased sensitivity to physical effects and the way such effects could influence their design resonates with the Animator's goal of encouraging novice animators to ``recreate life''. In addition to material jigs, future tools could explore the use of software jigs as have been introduced in woodworking~\cite{tian2021adroid}. Additionally, digital jigs such as a gyroscope, a buzzer, or an electromagnet could further shape the motion design experience, and may influence the puppeteer's analog methods.

While novice animators wouldn't be expected to generate polished animations during a first encounter with any novel animation tool, participants did deeply engage with the design process, and articulated perspectives on movement design that align with the goals of both the Animator and the Puppeteer. Even with the simple abstract model (see Figure ~\ref{fig:model-options}), participants immediately jumped up from the table, and used the provided materials to investigate different movement qualities. Participants did tend to anthropomorphize this simple geometric shape, probably a demonstration of the classic Heider and Simmel Illusion, where observers ascribe personality to moving geometric shapes~\cite{heider1944experimental}. As a next step, we hope to invite in participants who are more familiar with Blender, as well as expert animators to try the techniques and assess how such infleunces might fit into the professional's workflow.

Our findings position our tool as a low-fidelity, early prototype ``sketching'' style interface ideal for quickly generating a multiple options for movement, or for exploring a character's movement style. We imagine such tools used for exploratory animation work in tandem with established animation pipelines, which are already highly effective for precise control.


\katherine{one hole in the user study (besides N=2) that I think you need to address either here or in limitations is the idea that both of your participants are novices, but the story you build up before that point sort of hints at the idea that found object puppeteering could be of value to [professional] animators. Maybe you didn't actually intend to imply that, but the RW + interviews kind of led me to believe that.}

\rawn{so I think related to katherine's point is that I'm confused whether the focus of this paper is on making animation software easier to use for novices, making it more amenable to rapid prototyping, or leveraging the physical world with animation tools - to me the most compelling framing is the last one, and also I think is the one most supported by your evaluation. I think the argument might be more effective as 1. current animation tools, even for experts, are difficult to sketch in. 2. one argument for why they're difficult to sketch in is given by the puppeteer, who can easily sketch in his medium because he has the physical world as his material. 3. therefore, if we introduce the physical world into digital animation by incorporating some of the pupeteers practices, we can build an animation tool which is effective for sketching in. 4. In our user study we found this to be exactly the case - users preferred to use this interface for exploring the rough emotional 'outlines' of an animation in the sketching phase of the design process 5. we imagine these tools used for exploratory animation work in tandem with established animation pipelines, which are already highly effective for precise control.}

\section{Limitations and Future Work}
While this system requires the fairly extensive HTC Vive setup to use, we hope to inspire future designers to think about smaller, even more accessible tools that support the capturing, saving, collecting, or collaging of motion. Similar to the way many music artists keep collections of ``found'' audio clips (such as a dentists' drill, or the beep of a traffic light notification), we envision a broader engagement with movement as a design material. For example, imagine being able to quickly design the wave your Bitmoji does in a conversation with a friend. We imagine tools that support increased engagement with movement across many contexts: animation, application design, social media.

\section{Conclusion}

We have taken the first steps towards investigating the benefits of incorporating material objects as jigs into the animator's workflow. Our initial user study is an exploratory probe into the impact Found Object puppeteering techniques can have on novice animators as they engage in movement sketching. We end by celebrating the benefits of drawing on expertise from two separate but related fields, each with complimentary approaches.

\ep{feedback. Foreground movement sketching more in "authoring movement skethcing: inspiration from puppeteering in design". Include puppeteering ref in title. It's not how to support puppeteering. How do you sketch movement? Make sure it doesn't seem like an animation tool. "Movement sketching" sounds good for DIS. 

the puppetAR, that's the design of the architectured system that I built. Leads up to we built a system to evaluate how people might sketch these motions. It has components, we did a study, here's the report.

Experimental setup and how the hand controllers were used is shown well - great. Might include more info about where the control points on the puppets were. Have a picture w/ the rubber bands on arms to show more. People might be confused about Blender UI. No rendering / texturing. Highlight the control points on bones. Clarify what was used in the user study, why we used Blender. 

Fig. 3: Maybe take headset out of the picture entirely...

What are the essential elements of the system: 1) real-time 3D vis of articulated rigged models 2) a way for people to attach and manipulate up to 2 control points. Then, transition to specific details. Mention tracking. Could compare w/ fiducial tracking - need 6-DOF tracking, considered cameras (occlusion), IMU sensing would work --> VR controllers.

Share study details. Study isn't about movement accuracy, it's more about making it clear that this is about sketching / exploring a design space. It's about generating a diversity of movement, work w creative abstract concepts (low energy). Expand 4.2 description of study goals.

Could have open exploration in future work: hint at how people might save or iterate, collect the movements they generate. Motion could be exported into industrial tools, or kept as a collection of sketches used to drive movement in many contexts. [feels like a big jump for this paper]. 

Hinting at a design tool for designing motion. foley artists always sample the world, how could people always sample movement.

AR/VR world, games, social media.}

\revised{
As a future work, the calibration step of \systemOnesp could be improved by leveraging HoloLens's hand-tracking functionality.
The user could stare at the hand to get a reading of the hand's position in HoloLens's coordinate system.
Repeating this step at various locations, we could get several pairs $(x_i, x_i')$, each satisfying
$$x_i' = kAx_i + b$$
Then we could solve $kA$ and $b$ with least squares.
Furthermore, we could readjust the matching with more readings during the authoring process.

We designed our interface to have a `menu-selection' interface similar to Blender's specifically to keep the overall design closer to Blender's style and allow us to compare the two. However, this limitation undermined some of the advantages of such an immersive environment.
} 

\balance
\bibliographystyle{ACM-Reference-Format}
\bibliography{puppet}

\revised{
\subsection{Animator techniques and tools}
Character Concept Artists for Games~\url{https://www.youtube.com/watch?v=lA5HG8Q4sKg}. Panelist 1 talked about how putting characters in context (``cinematic context'') is a better way to communicate with stakeholders (rather than showing a lineup with just appearance as relevant info). Panelist 2 talked about how char design would affect movement, animation, etc. Shouldn't that go both ways? 

\subsection{Creative tools for creative practices}
Prior work with dancers has shown the value of specific tools for early on in the design process~\cite{ciolfi2018knotation}.

\section{Links}
The 7 Workflows of Professional Animators by Sir Wade
\url{https://www.youtube.com/watch?v=GMTet6nd_iM&ab_channel=SirWadeNeistadt}

The Secret Animation Workflow You Should Be Using
\url{https://www.youtube.com/watch?v=7dAzk2oeQoA&ab_channel=SirWadeNeistadt}

Pixar livestream
\url{https://www.instagram.com/tv/CMcm4r_L5Uw/}

Blender - Making an Animated Short Film, start at ~44:36
\url{https://www.youtube.com/watch?v=0knhZcEyhh4&t=2397s&ab_channel=Markom3D}

Arduino to Blender Tutorial: Control Virtual Worlds with the Real One
\url{https://www.youtube.com/watch?v=CypV9pPTCXo}

Dragon model (rigged)
\url{https://www.cgtrader.com/free-3d-models/animals/reptile/dragon-version-1-rigged-and-game-ready}

Gesture recognition for interactive prototypes.
\url{http://depts.washington.edu/acelab/proj/dollar/impact.html}

\subsection{IMU / Pose estimation stuff}
Using Inertial Sensors for Position and Orientation Estimation by Kok et al. \url{https://arxiv.org/pdf/1704.06053.pdf}

Discussion on getting position from MPU6050 ~\url{https://www.researchgate.net/post/What_is_the_best_way_to_measure_position_with_an_accelerometer_and_gyroscope}

Pose estimation lecture from Stanford ~\url{https://stanford.edu/class/ee267/lectures/lecture11.pdf}

StackOverflow description on how to do pose estimation on arduino~\url{https://robotics.stackexchange.com/questions/16757/what-is-the-algorithm-to-get-position-linear-displacement-and-linear-velocity}

Code sample to calculate walking.~\url{https://conservancy.umn.edu/bitstream/handle/11299/199815/Real-Time\%20Position\%20Tracking\%20Using\%20IMU\%20Data.pdf?sequence=1&isAllowed=y}

Discussion around the problems calculating position from an IMU:~\url{https://web.archive.org/web/20201111202735/http://www.chrobotics.com/library/accel-position-velocity} Commercial IMUs (price unlisted): ~\url{https://inertiallabs.com/products/imup/}

mpu6050 on a quadcopter~\url{http://iieng.org/images/proceedings_pdf/AE0516306.pdf}

OpenXR on Blender, for Vive and maybe also Oculus?~\url{https://blender.community/c/rightclickselect/hLfbbc/?sorting=hot} ~\url{https://blenderartists.org/t/openvr-tracker-streaming/1236428}. But it uses PyOpenVR which I think has entered maintenance mode~\url{https://github.com/cmbruns/pyopenvr/releases}. Here's the plugin~\url{https://blenderartists.org/t/openvr-tracker-streaming/1236428}. 

PyOpenXR (new draft vs PyOpenVR): \url{https://github.com/cmbruns/pyopenxr}

Here's another BlenderXR project but it wasn't updated since July 2020~\url{https://www.marui-plugin.com/blender-xr/}

\section{Installation steps}

\subsection{MarUI}
\begin{itemize}
    \item Instructions for working w/ the mouse while a Blender VR window is open: \url{https://github.com/MARUI-PlugIn/BlenderXR/issues/12}
    \item Patch to use Blender VR w/o HMD \url{https://github.com/MARUI-PlugIn/BlenderXR/issues/18}
    \item Discussion about using controller to pose a char \url{https://github.com/MARUI-PlugIn/BlenderXR/issues/20}
    \item Discussion about build errors with blender: \url{https://devtalkx.blender.org/t/make-update-error-required-libraries-not-found/11859}
    \item MarUI blender debug version tries to install this link first: \url{https://svn.blender.org/svnroot/bf-blender/trunk/lib/win64_vc14}, which fails. THe parent folder there only contains vc15, not 14, which luckily exists. Just need to update the \url{check_libraries.cmd} file to get those installed
    \item The actual tagged release for blender 2.82 (the version released w MarUI) lists both 14 and 15: \url{https://svn.blender.org/svnroot/bf-blender/tags/blender-2.82-release/lib/}
    \item Autokeying is not supported yet (as of Dec 19) \url{https://github.com/MARUI-PlugIn/BlenderXR/issues/35}
    \item Updated all refs to v14 to v15, and python path from 37 to 39 (in format.cmd and find-dependencies.cmd, nuke.cmd, \url{make_update.py}, \url{platform_win32.cmake}, format.cmd, and "source/creator/CMakeLists.txt".
\end{itemize}

\subsection{steve's plugin}
\begin{itemize}
    \item install gitbash.
    \item Install latest version of Python 3.9.7, which installs pip. Make sure to select "install to PATH" when installing on Windows. Confirm python installation by opening a Gitbash window somewhere and typing `python --version''
    \item Download the latest version of Blender (I'm using 2.93 - LTS). (if you want to hack Blender, don't use the version from the store, use the portable .zip \url{https://www.blender.org/download/release/Blender2.93/blender-2.93.6-windows-x64.zip/?x71978}. Note that the below is technically a Blender hack, so you must use the portalbe .zip).
    \item Use pip to install PyOpenVR: `pip install openvr` ~\url{https://github.com/cmbruns/pyopenvr/releases}. If pip doesn't tell you where pyopenvr was installed, try installing it again and it'll show you where. 
    \item Move PyOpenVR files installed via pip to Blender python folder. They were here \url{C:\\Users\\molly\\AppData\\Local\\Programs\\Python\\Python39\\Lib\\site-packages} and I moved them here: \url{C:\\Program Files\\Blender Foundation\\Blender 2.93\\2.93\\python}
    \item Then, in the addons part of Blender (Edit -> Preferences), enable 3D Navigation.
    \item Add the plugin.
    \item Open the 3D navigation sidebar (see Figure blendervr-instructions), then open the OpenVR streaming tab.
\end{itemize}

FOR QUEST
\begin{itemize}
    \item Setup Steam VR to work with Oculus Quest \url{https://www.youtube.com/watch?v=n5fWoYi5100&ab_channel=BMFVR}
\end{itemize}

CURRENT STATUS
trying to print out from new openvr blender add-on, so I can see why Streaming from Oculus Quest doesn't work.

\subsection{RIGGING IN BLENDER}
Making a char that's aligned right.
\begin{itemize}
    \item build char.
    \item select a view. (i.e., the front view). open Edit mode.
    \item A to select all the bones.
    \item shift+N to recalculate roll. Select "view axis" - all the bones should rotate properly towards the view. \url{https://www.youtube.com/watch?v=jp_SqjB0468&ab_channel=SebastianLague}
    \item done.
    \item to test parenting: in Pose mode, grab a bone, and make sure all the "downstream' bones move with it. \url{https://www.youtube.com/watch?v=jp_SqjB0468&ab_channel=SebastianLague}
    \item Next, \textbf{you need to align the axes} \url{https://www.youtube.com/watch?v=suP14lYWpN8&ab_channel=LevelPixelLevel}
\end{itemize}

Adding control bones.
\begin{itemize}
    \item Build your char.
    \item Add control bones.
    \item Clear parent of control bones. (disconnecting seems irrelevant? Neither harms nor helps...)
    \item Switch to Pose mode. Select the bone you want to move. Add IK constraint to it. Target is Armature, Bone is the control bone.
    \item To add rotation: Edit mode. Select the bone you want to rotate. Shift-select the control bone. Ctrl-P (keep offset). Next add a "copy location" constraint. 
    \item To add a pole target: extrude bone where you want it. Alt-P to disconnect from hierarchy.
\end{itemize}

\subsection{Rigging tutorials}
\begin{itemize}
    \item Youtube tutorial to make the li'l doggy (non-humanoid-v2): IK? | Blender Inverse Kinematics | Beginners Tutorial - \url{https://www.youtube.com/watch?v=nkNN4aJjjYg}
    \item Vague memory of this one being good: Full Robot Rig - Introduction | Rigging For Animation \url{https://www.youtube.com/watch?v=GKvJKV5fiIU}. The rest of the series might be important \url{https://www.youtube.com/watch?v=sugBFqZ-qtY}, \url{https://www.youtube.com/watch?v=OH2CczT8v7Q}, \url{https://www.youtube.com/watch?v=P3_zi-YcWfU&t=419s}, 
    \item I used this one a lot, but I think in the end I needed a combo of this one and the li'l doggy one. Easy Character Rigging with Inverse Kinematics | Blender 2.92 | Beginners tutorial \url{https://www.youtube.com/watch?v=hWfUe03Ib5E&}
    \item RPG graphics E01: Character model [Blender] \url{https://www.youtube.com/watch?v=aAO4C_8y0w8}
    \item Can't remember if this one was useful: RPG graphics E02: Character rig [Blender] \url{https://www.youtube.com/watch?v=jp_SqjB0468&t}
    \item Can't remember: How To Control The Pole Target - Blender IK Leg Rig Tutorial\url{https://www.youtube.com/watch?v=suP14lYWpN8}
    \item Blender 2.91 | Human Meta Rig | Rigify (Easy Beginners) \url{https://www.youtube.com/watch?v=tzJ39ZOhfQ4&}
    \item Armatures - Blender 2.80 Fundamentals - I think this one and the related file was actively harmful \url{https://www.youtube.com/watch?v=cZ3o5tjO51s}
    \item Blender 2.8 Tutorial : Rig ANY Character for Animation in 10 Minutes! \url{https://www.youtube.com/watch?v=SBYb1YmaOMY&t=337s}
    
\end{itemize}

\section{Design}
\begin{itemize}
    \item Passive augmentation (weights, stretchy band, plastic bars, magnet, etc)
    \item Active augmentation (gyroscope, buzzer, etc)
    \item Change digital mapping - require more movement to get less effects on screen. Random variation as a warm-up exercise.
\end{itemize}

\url{https://www.youtube.com/watch?v=lA5HG8Q4sKg} Panelist 2 in this Game Character Design talk gives four elements of the setting/universe to consider: time period/setting (old-timey, modern, futuristic); realistic vs stylized; literal vs symbolic; playful vs serious. Should any of these shape the way I explore the character design space?

\section{Phrases I like}

For example, we can frame her use of Mode Switching as `knowing-through-action': combining her expertise as a professional performer with the tools of video-recording and writing together produces output -- in this case, a scene -- that is meaningful and that moves her design process forward. This `knowing-through-action' arises as she leverages different tools throughout her process. She explicitly describes the different `modes of work' she taps into by using the video-camera, or the journal, and how these tools then shape the mindset she has and the way she interacts with her own output.  
Distributed cognition presents a similar lens for understanding this concept, which also embraces the larger context of her working environment as part of her cognition. In other words, her creative process is an emergent property of the interaction between her own skills and the camera or the journal. 

Seen through the lens of \textit{instruments of inquiry} or distributed cognition, we can see the importance of understanding how closely enmeshed the creative behavior is with the tools at hand.  In the example described above, switching modalities (from video-taping to writing) was nearly synonymous with switching creative modes (from generating to editing). It is difficult to separate the thinking and doing aspects of her working style, and difficult to separate the goal of the task from the tools used in that task. How Bodies Matter~\cite{klemmer2006bodies} showed the motivation for tangible tools -- here we see participants using digital tools in particular roles to mode-switch between different parts of a creative task.  This motivates the design of tools with distinctive and memorable interfaces, tools that take advantage of different modalities, and even encourage physical interactions to take advantage of muscle memory when appropriate.

Whether designing for human robot interaction, mixed reality, social media, or games, designers engaging in all forms of creative work would benefit from access to CSTs that enable the rich, embodied design experiences. My research understands the creative process as a form of “knowing-through-action”, or the construction of new knowledge generated through action carried out together with a computational tool. This understanding closely parallels Dalsgaard’s notion of instruments of inquiry, which understands the creative process as "intertwined" and "co-evolving with" the environment and tools[12]. Distributed cognition presents a similar lens for understanding this concept; both see the creative process as an emergent property of the interaction between her own skills and the tools or system in use[13].

\section{Framing options}
Looking at the creative process of character design in depth, and introducing 3 different tools to support this early ideation: motion path planning, character movement design, and a reference library. Supporting evidence: in-depth formative interview w/ professional animator.  

Specifically talking about early char movement, ideation process. Focus on need to perturb the process, and how the way that puppeteers do that could be a useful blueprint for such a process. 

If we did the 2nd one, is there a way to shoe-horn in the tangible choreo project? Or maybe just talk about them as two separate aspects of something similar: motion path planning and movement design? Or just gie up on that. I think I"ll abandon that for now.

\section{Dec 7 Intro}
When designing a new character, or creating the movement for a particular scene, animators seek out or create references. These short video clips act as inspiration and example, providing a naturalistic framework for the final design.

\section{2021-09-01 Brain dump}
Just like how tick-tock doesn't let you edit anything that you filmed, because that would make you hyper focused on creating the perfect video instead generating many different videos. This system is meant to support quick, low fidelity, rapid iteration. A tool for thinking. A sketch interface.

Meant to support early character design. 

Import philosophies, approaches, techniques, patterns, etc from puppeteering into 3D animation. Perturb the thinking of 3D animators with expertise from tangible puppeteers.

Exploring the benefit of tangible experiences over digital ones in this scenario. 

Passive additions (weight, stretchy band, stick) and active ones: gyroscope, buzzer, etc.

Puppets: photorealistic, cartoon, abstract humanoid, fully abstract

Movement as a material. Ways to effect movement: Weight, timing.

If Baby Yoda was controlled by more than one person, can use that as a reason to motivate this design. First step towards paired, remote-controlled digital/physical puppet systems.

} 

\end{document}

%% file: macros.tex
\newcommand\systemOne{{ChoreographAR}}

\newcommand\systemOnesp{\systemOne\ }

\newcommand\systemTwo{{PuppetJig}}

\newcommand\systemTwosp{\systemTwo\ }

\newtoggle{comments}
\toggletrue{comments}
\togglefalse{comments}

\iftoggle{comments} {
  \newcommand {\katherine}[1]{{\color{blue}\bf{KAT: #1}\normalfont}}
  \newcommand {\ep}[1]{{\color{cyan}\bf{EP: #1}\normalfont}}
  \newcommand {\molly}[1]{{\color{orange}\bf{MN: #1}\normalfont}}
  \newcommand {\rawn}[1]{{\color{magenta}\bf{ER: #1}\normalfont}}
  \newcommand {\sarah}[1]{{\color{red}\bf{SS: #1}\normalfont}}
  \newcommand {\rut}[1]{{\color{purple}\bf{rt: #1}\normalfont}}
  \newcommand {\jessie}[1]{{\color{cyan}\bf{JM: #1}\normalfont}}
  \newcommand {\andrew}[1]{{\color{cyan}\bf{AH: #1}\normalfont}}
  \newcommand {\nate}[1]{{\color{cyan}\bf{NW: #1}\normalfont}}
  \newcommand {\tha}[1]{{\color{cyan}\bf{THA: #1}\normalfont}}

}{
  \newcommand {\katherine}[1]{}
  \newcommand {\ep}[1]{}
  \newcommand {\molly}[1]{}
  \newcommand {\christie}[1]{}
  \newcommand {\sarah}[1]{}
  \newcommand {\rut}[1]{}
  \newcommand {\rawn}[1]{}
  \newcommand {\jessie}[1]{}
  \newcommand {\andrew}[1]{}
  \newcommand {\nate}[1]{}
  \newcommand {\tha}[1]{}
 }

\newenvironment{myquote}{\list{}{\leftmargin=0.02\textwidth \rightmargin=0.02\textwidth}\item[]}{\endlist}
\newcommand*{\participant}[1]{{\textit{\small{\fontfamily{cmss}\selectfont{#1}}}}:}
\newcommand*{\quoted}[1]{{\small{\fontfamily{cmss}\selectfont{#1}}}}
\newcommand{\squote}[2]{\begin{myquote}\quoted{\participant{#1} #2}\end{myquote}}

\newtoggle{showRevised}
\toggletrue{showRevised}
\togglefalse{showRevised}

\iftoggle{showRevised} {
\newcommand {\revised}[1]{{\color{blue} #1}}
}{
\newcommand {\revised}[1]{}
}